\documentclass{appolb}
\usepackage{graphicx}
\usepackage{combelow}
\usepackage{amsmath}
\usepackage{amssymb}
\usepackage{bm}
\usepackage{slashed}
\usepackage{braket}

\def\tr{{\text{tr}}}

\def\vx{\mathbf{x}}
\def\hatt{{\hat{t}}}
\def\hati{{\hat{\imath}}}
\def\hatz{{\hat{z}}}
\def\halpha{{\hat{\alpha}}}

\def\hsigma{{\hat{\sigma}}}

\def\vgamma{\bm{\gamma}}

\begin{document}
\title{Fermion condensation under rotation on \\
anti-de Sitter space 
\thanks{Presented at The 6th Conference of the Polish Society on Relativity,
Szczecin, Poland, 23-26.09.2019}%
}
\author{Victor E. Ambru\cb{s}
\address{Department of Physics, West University of Timi\cb{s}oara, Bd. Vasile P\^arvan 4, Timi\cb{s}oara 300223, Romania}
}
\maketitle
\begin{abstract}
Due to the local curvature, the fermion condensate (FC) for a free Dirac 
field on anti-de Sitter (adS) space becomes finite, even in the massless limit.
Employing the point splitting method using an exact expression for the Feynman 
two-point function, an experssion for the local FC is derived. Integrating 
this expression, we report the total FC in the adS volume and on its boundary.
\end{abstract}
% \PACS{PACS numbers come here}
  
\section{Introduction}\label{sec:intro}

Over the past couple of decades, the analysis of quantum field theory (QFT) 
on the anti-de Sitter (adS) background space-time has received much attention 
due to the conjectured adS/CFT correspondence \cite{aharony2000}. Through 
this conjecture, important insight into the properties of the quark-gluon 
plasma formed in relativistic heavy-ion collisions was drawn \cite{kovtun2005}.

Recent experiments performed by the STAR collaboration revealed the 
polarisation of the QGP in non-central collisions \cite{STAR2017}.
One mechanism that could lead to this polarisation is the chiral vortical 
effect, due to the spin-orbit coupling predicted through the Dirac 
equation \cite{kharzeev16}.

In this contribution, we present a study of thermal states of fermions undergoing 
rigid rotation on the anti-de Sitter space. The focus of this study is the fermion 
condensate (FC) induced by the coupling to curvature. The discussion is restricted 
to massless particles in the absence of interaction. 

\section{Finite temperature expectation values}\label{sec:qft}

The line element of adS can be written as:
\begin{equation}
 ds^2 = \frac{1}{\cos^2\omega r}
 \left[-dt^2 + dr^2 + \frac{\sin^2\omega r}{\omega^2} \left(d\theta^2 + \sin^2\theta d\varphi^2\right)\right],
\end{equation}
where $t \in (-\infty, \infty)$,\footnote{We consider the covering space of adS.}
$0 \le \omega r < \frac{\pi}{2}$ and 
the inverse radius of curvature $\omega$ is related to the Ricci scalar through
$R = -12\omega^2$. We further employ the following
Cartesian gauge tetrad \cite{cotaescu2007}:
\begin{align}
 e_{\hatt} =& \cos \omega r \, \partial_t, &
 e_{\hati} =& \cos \omega r \left[ \frac{\omega r}{\sin \omega r} \left(
 \delta_{ij} - \frac{x^ix^j}{r^2}\right) + \frac{x^ix^j}{r^2}\right] \partial_j,
%  \nonumber\\
%  \omega^\hatt =& \frac{dt}{\cos\omega r}, &
%  \omega^{\hati} =& \frac{1}{\cos\omega r}\left[\frac{\sin \omega r}{\omega r}
%  \left(\delta_{ij} - \frac{x^ix^j}{r^2}\right) + \frac{x^ix^j}{r^2}\right]dx^j. 
 \label{eq:frame}
\end{align}
by which the local gamma matrices $\gamma^\mu = e^\mu_\halpha \gamma^\halpha$ 
are written in terms of the Minkowski ones, which satisfy 
$\{\gamma^\halpha, \gamma^\hsigma\} = -2\eta^{\halpha\hsigma}$.
At finite temperature $\beta_0^{-1}$ and in rigid rotation with angular velocity 
$\bm{\Omega} = \Omega \bm{k}$, we have \cite{vilenkin80}
\begin{equation}
 \braket{\hat{\overline{\Psi}}\hat{\Psi}}_{\beta_0,\Omega} = \mathcal{Z}^{-1} {\rm tr}(\hat{\rho} 
 \hat{\overline{\Psi}}\hat{\Psi}), \qquad \hat{\rho} = e^{-\beta_0(\hat{H} - \Omega \hat{M}^\hatz)},
 \label{eq:tev_def}
\end{equation}
%In the above, $::$ denotes Wick ordering, 
where $\hat{H} = i \partial_t$,
$\hat{M}^\hatz = -i\partial_\varphi + S^\hatz$, 
$S^\hatz = \frac{i}{2} \gamma^{\hat{1}} \gamma^{\hat{2}}$
and $\mathcal{Z} = {\rm tr}(\hat{\rho})$.

To evaluate Eq.~\eqref{eq:tev_def}, we take the point-splitting approach, by which 
\cite{groves2002}
\begin{equation}
 \braket{\hat{\overline{\Psi}}\hat{\Psi}}_{\beta_0,\Omega} = 
 -\lim_{x'\rightarrow x} \tr [iS^F_{\beta_0,\Omega}(x,x') \Lambda(x',x)],
 \label{eq:tev}
\end{equation}
% where $\Delta S^F_{\beta_0, \Omega}(x,x') = S^F_{\beta_0, \Omega}(x,x') - S^F_{\rm vac}(x,x')$ 
% is the difference between the thermal two-point function, 
where $S^F_{\beta_0, \Omega}(x,x')$ is the thermal two-point function and $\Lambda(x,x')$ is the 
bispinor of parallel transport, given by \cite{ambrus2017}:
\begin{multline}
 \Lambda(x,x') = \frac{\sec(\omega s/2)}{\sqrt{\cos\omega r \cos\omega r'}} \Bigg[\\
 \cos\frac{\omega \Delta t}{2} \left(
 \cos\frac{\omega r}{2} \cos\frac{\omega r'}{2} +
 \sin\frac{\omega r}{2} \sin\frac{\omega r'}{2}
 \frac{\vx \cdot \vgamma}{r} \frac{\vx' \cdot \vgamma}{r'}\right)\\
 + \sin\frac{\omega \Delta t}{2} \left(
 \sin\frac{\omega r}{2} \cos\frac{\omega r'}{2} \frac{\vx \cdot \vgamma}{r} \gamma^\hatt
 +\sin\frac{\omega r'}{2} \cos\frac{\omega r}{2} \frac{\vx' \cdot \vgamma}{r'} \gamma^\hatt\right)\Bigg].
\end{multline}
Using the property 
%\begin{equation}
$\hat{\rho} \hat{\Psi}(t,\varphi) \hat{\rho}^{-1} = 
 e^{-\beta_0 \Omega S^\hatz} \hat{\Psi}(t + i \beta_0, \varphi + i \beta_0 \Omega)$,
%\end{equation}
together with the imaginary time anti-periodicity of the two-point function \cite{laine2016}, 
% Using a canonical formalism \cite{birrell1982}, 
it is possible to compute $S^F_{\beta_0,\Omega}(\Delta t, \Delta \varphi) \equiv S^F_{\beta_0,\Omega}(t, \varphi; t', \varphi')$ 
via \cite{birrell1982}:
\begin{equation}
 S^F_{\beta_0,\Omega}(\Delta t, \Delta \varphi) = \sum_{j = -\infty}^\infty (-1)^j 
 e^{-j \beta_0 \Omega S^\hatz} S^F_{\rm vac}(\Delta t + ij\beta_0, \Delta \varphi + i j \beta_0 \Omega),
\end{equation}
where $S^F_{\rm vac}(x,x')$ is the vacuum two-point function.
% , while $e^{a S^\hatz} = \cosh \frac{a}{2} + 2a S^\hatz \sinh\frac{a}{2}$. 
The above expression is valid only when the vacua ($\beta_0 \rightarrow 0$) corresponding to the rotating 
(fixed $\Omega$) and non-rotating ($\Omega = 0$) cases coincide. This is ensured on adS when $|\Omega| \le \omega$ \cite{ambrus2014}, which we assume to hold
for the remainder of this paper.

Due to the maximal symmetry of adS, $S_{\rm vac}^F(x,x')$ can be written as \cite{muck}:
% propagator can be written as:
\begin{equation}
 iS_{\rm vac}^F(x,x') = [{\mathcal {A}}(s) + {\mathcal {B}}(s) \slashed{n}] \Lambda(x,x'),
\end{equation}
where $n_\mu = \nabla_{\mu} s(x,x')$ is the normalised tangent at $x$ to the geodesic connecting 
$x$ and $x'$, while the geodesic interval $s$ is given through:
\begin{align}
 \cos \omega s =& \frac{\cos\omega \Delta t}{\cos\omega r \cos \omega r'} - 
 \cos\gamma \tan \omega r \tan \omega r', 
\end{align}
where $\gamma$ is the angle between $\bm{x}$ and $\bm{x}'$, such that 
$\cos\gamma = \cos\theta \cos \theta' + \sin\theta \sin \theta' \cos\Delta\varphi$.
For massless fermions, 
the functions $\mathcal{A}$ and $\mathcal{B}$ are 
\begin{equation}
 {\mathcal {A}}\rfloor_{M = 0} = \frac{\omega^3}{16\pi^2} \left(\cos\frac{\omega s}{2}\right)^{-3}, \qquad
 {\mathcal {B}}\rfloor_{M = 0} = \frac{i\omega^3}{16\pi^2} \left(\sin\frac{\omega s}{2}\right)^{-3}. 
\end{equation}
% The bi-spinor of parallel transport is given by 

\section{Analysis and conclusions}

Without presenting the details of the computation, we find \cite{ambrus2014}
\begin{equation}
 \braket{:\hat{\overline{\Psi}}\hat{\Psi}:}_{\beta_0,\Omega} = 
 \sum_{j = 1}^\infty \frac{(-1)^{j+1} (\cos\omega r)^4 \cosh \frac{\omega j \beta_0}{2} \cosh \frac{\Omega j \beta_0}{2}}
 {2\pi^2 [\sinh^2(\frac{\omega j \beta_0}{2}) + \cos^2 \omega r - \sin^2 \omega r \sin^2 \theta \sinh^2(\frac{\Omega j \beta_0}{2})]^2}.
 \label{eq:FC}
\end{equation}
The total FC can be obtained by integrating Eq.~\eqref{eq:FC} over the whole space:
\begin{multline}
 V^{\rm FC}_{\beta_0, \Omega} = \int d^3x \sqrt{-g} \braket{:\hat{\overline{\Psi}}\hat{\Psi}:}_{\beta_0,\Omega} =
 -\sum_{j = 1}^\infty \frac{(-1)^j \cosh(\frac{\Omega j \beta_0}{2}) / \sinh(\frac{\omega j \beta_0}{2})}
 {\cosh(\omega j \beta_0) - \cosh(\Omega j \beta_0)} \\
 \simeq \frac{3 \zeta(3) T_0^3}{\omega(\omega^2 - \Omega^2)} - 
 \frac{(3\omega^2 - \Omega^2) T_0}{6 \omega(\omega^2 - \Omega^2)} \ln 2 + O(T_0^{-1}).
 \label{eq:V}
\end{multline}
\begin{figure}[htb]
\centerline{%
\begin{tabular}{cc}
 \includegraphics[width=0.45\linewidth]{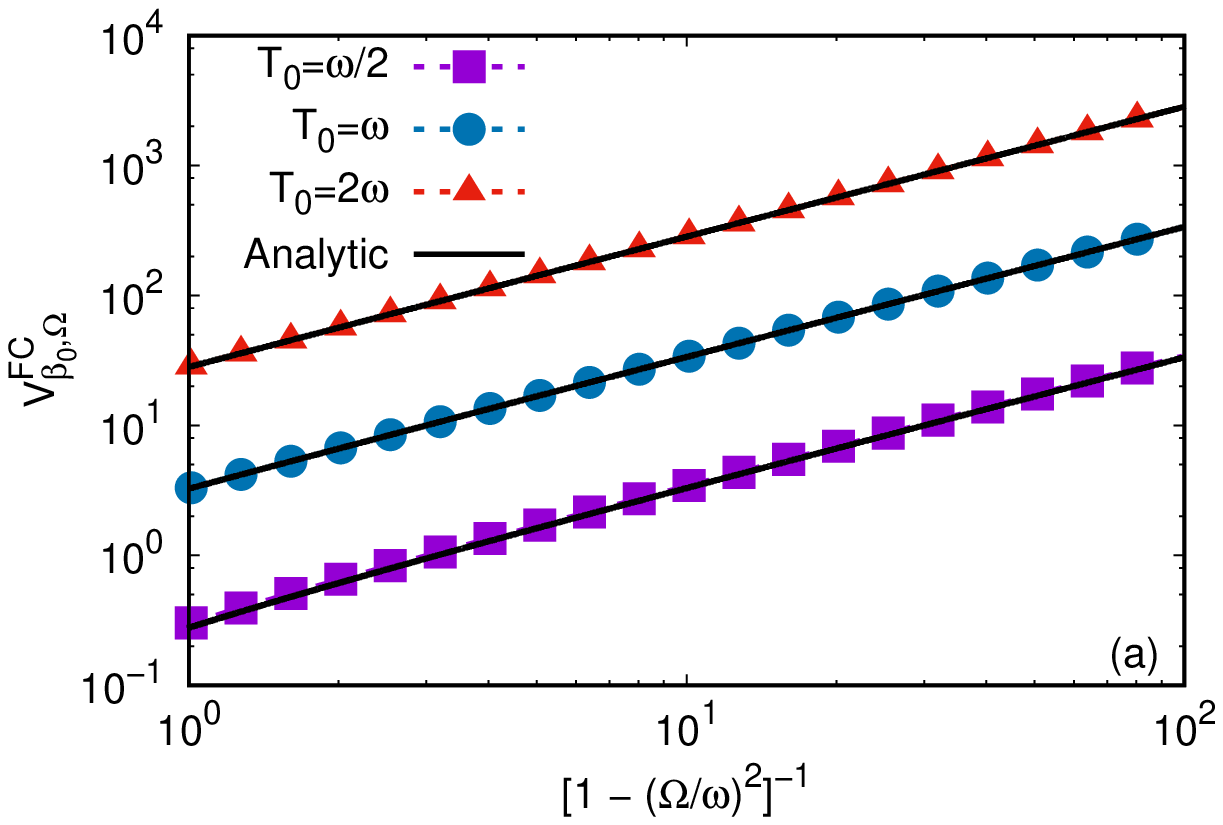} &
 \includegraphics[width=0.45\linewidth]{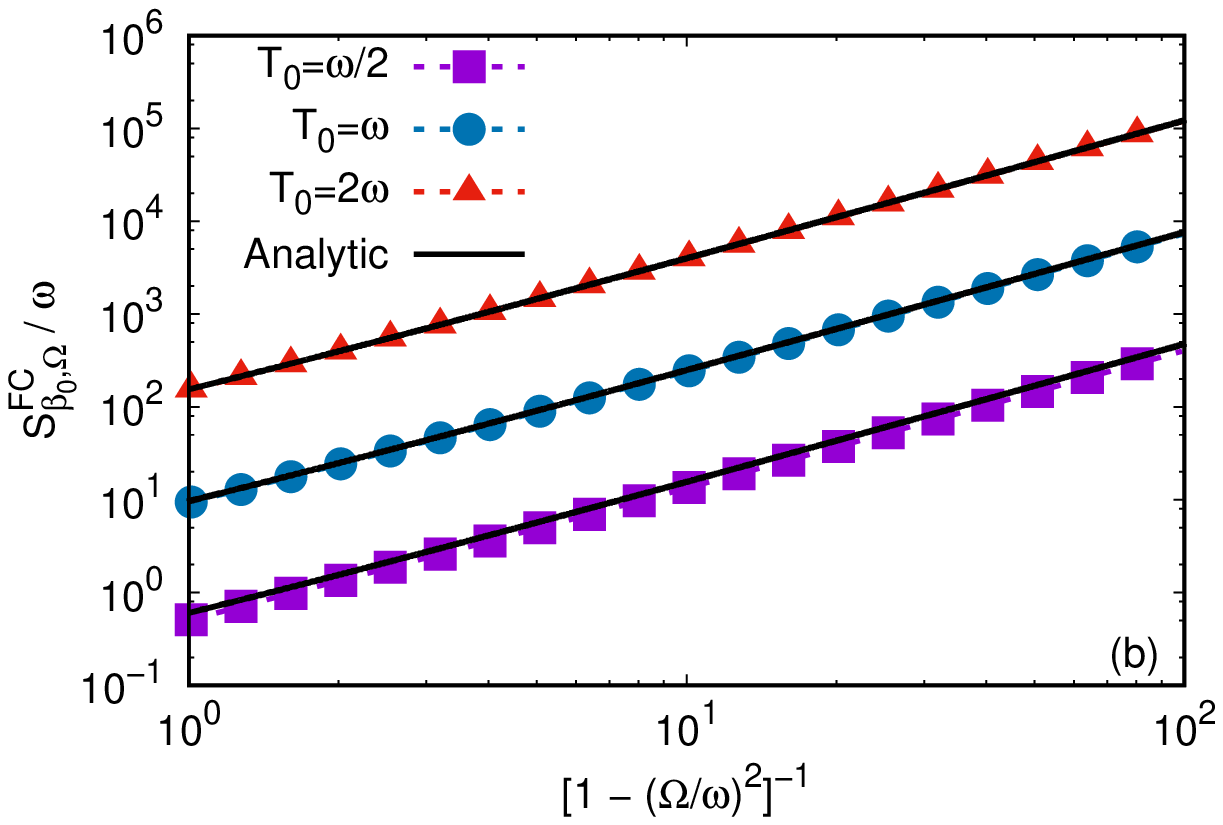}
\end{tabular}
}
\caption{Dependence of (a) $V^{\rm FC}_{\beta_0, \Omega}$ and (b) $S^{\rm FC}_{\beta_0, \Omega} / \omega$ with 
respect to $(1 - \Omega^2 / \omega^2)^{-1}$, in logarithmic scale. The dotted lines and 
symbols are numerical results obtained using Eq.~\eqref{eq:FC}, while the 
analytic curves correspond to Eqs.~\eqref{eq:V} and \eqref{eq:S}.}
\label{fig:res}
\end{figure}
On the boundary, the following result is obtained:
\begin{multline}
 S^{\rm FC}_{\beta_0,\Omega} = \int d\Omega \sqrt{-g} \braket{:\hat{\overline{\Psi}}\hat{\Psi}:}_{\beta_0,\Omega} \\
%  =
%  -\frac{\omega}{\pi} \sum_{j = 1}^\infty \frac{(-1)^j \cosh(\frac{\omega j \beta}{2}) \cosh(\frac{\Omega j \beta_0}{2})}
%  {[\cosh^2(\omega j \beta_0) - \cosh^2(\Omega j \beta_0)]^2}\\
%  \times \left[\frac{1}{\sinh^2(\frac{\omega j \beta_0}{2})} + 
%  \frac{\arctan[\sinh(\frac{\Omega j \beta_0}{2}) / \sqrt{\cosh^2(\frac{\omega j \beta_0}{2}) - \cosh^2(\frac{\Omega j \beta_0}{2})}]}
%  {\sinh(\frac{\Omega j \beta_0}{2}) \sqrt{\cosh^2(\frac{\omega j \beta_0}{2}) - \cosh^2(\frac{\Omega j \beta_0}{2})}}
%  \right] \\
 \simeq \frac{7\pi^3 T^4}{45(\omega^2 - \Omega^2)^{3/2}} 
 \left[\frac{\omega}{\Omega} {\rm tan}^{-1}\left(\frac{\Omega / \omega}{\sqrt{1 - \frac{\Omega^2}{\omega^2}}}\right) + \sqrt{1 - \frac{\Omega^2}{\omega^2}}\right] + O(T^2).
 \label{eq:S}
%  \nonumber\\
%  \simeq& \frac{7\pi^3 T^4}{45\omega(\omega^2 - \Omega^2)}  
%  \left\{1 + \frac{2\Omega^2}{3\omega^2} + O\left[\left(\frac{\Omega}{\omega}\right)^4\right]\right\}.
\end{multline}
Both $V^{\rm FC}_{\beta_0,\Omega}$ \eqref{eq:V} and $S^{\rm FC}_{\beta_0,\Omega}$ are amplified 
due to the rotation through the prefactors $(1 - \Omega^2 / \omega^2)^{-1}$ and 
$(1 - \Omega^2 / \omega^2)^{-3/2}$, respectively.
Figs.~\ref{fig:res}(a) and \ref{fig:res}(b) show the dependence of 
$V^{\rm FC}_{\beta_0,\Omega}$ and $S^{\rm FC}_{\beta_0,\Omega}$ on $(1 - \Omega^2 / \omega^2)^{-1}$, for 
various values of the temperature $T_0 = \beta_0^{-1}$. It can be seen that the 
analytic results \eqref{eq:V} and \eqref{eq:S} (shown with solid black lines) match
well the numerical results (dotted lines and symbols) computed using Eq.~\eqref{eq:FC}.

{\bf Acknowledgments.}
This work was supported by a Grant from the Romanian 
National Authority for Scientific Research and Innovation, CNCS-UEFISCDI,
project number PN-III-P1-1.1-PD-2016-1423.

\end{document}